\documentclass[a4paper,12pt]{article}
\usepackage[utf8]{inputenc}
\usepackage{cancel}
\usepackage{tikz}
\usepackage{ulem}
\usepackage{amsfonts}
\usepackage{amssymb}
\usepackage{graphicx}
\usepackage{amsmath}
\usepackage{enumerate}
\usepackage{mathtools}
\usepackage{subfig}
\usepackage{color}
\usepackage{tikz}
\usepackage{float}
\usepackage{bbm}
\usepackage{here}
\usepackage{cite}
\usepackage{mathrsfs}
\usepackage{float,epsfig}
\usepackage{dcolumn}
\usepackage{graphicx}
\usepackage{bm}
\usepackage{amsmath,amssymb,amsthm}
\usepackage[colorlinks=true,linkcolor=blue,citecolor=red]{hyperref}
\usepackage{multirow}
\usepackage[toc,page]{appendix}
\usetikzlibrary{arrows.meta}
\usetikzlibrary{bending}
\usetikzlibrary{calc}
\newcommand{\be}{\begin{equation}}
\newcommand{\ee}{\end{equation}}
\newcommand{\bea}{\setlength\arraycolsep{2pt} \begin{eqnarray}}
\newcommand{\eea}{\end{eqnarray}}

\def\0{{\sst{(0)}}}
\def\1{{\sst{(1)}}}
\def\2{{\sst{(2)}}}
\def\3{{\sst{(3)}}}
\def\4{{\sst{(4)}}}
\def\5{{\sst{(5)}}}
\def\6{{\sst{(6)}}}
\def\7{{\sst{(7)}}}
\def\8{{\sst{(8)}}}
\def\sst#1{{\scriptscriptstyle #1}}

\makeatletter \@addtoreset{equation}{section}

\definecolor{lime}{HTML}{A6CE39}

\begin{document}

\title{{\normalsize \textbf{\Large On Stability    Behaviors of  5D   M-theory Black Objects     }}}
\author{ {\small Adil  Belhaj\footnote{a-belhaj@um5r.ac.ma}\; and  Abderrahim  Bouhouch\footnote{abderrahim.bouhouch@um5r.ac.ma }    \thanks{\bf  Authors in alphabetical order.} \hspace*{-8pt}} \\
%EndAName
{\small D\'{e}partement de Physique, \'Equipe des Sciences de la
mati\`ere et du rayonnement, ESMaR} \\ {\small Facult\'e des Sciences, Universit\'e Mohammed V de Rabat, Rabat,
Morocco}}
\maketitle

\begin{abstract}
Using    $N = 2$  supergravity formalism,   we   investigate certain behaviors  of five dimensional  black    objects      from  the  compactification of M-theory 
on  	 a  Calabi-Yau three-fold.   The manifold    has been constructed as    the intersection of two homogeneous polynomials of degrees $  (\omega+2,1)$ and $ (2,1) $ in  a product of  two weighted projective spaces given by  $ \mathbb{WP}^{4}(\omega,1,1,1,1) \times\mathbb{P}^{1}$.     First, we determine the allowed electric charge  regions  of the  BPS  and non BPS  black holes   obtained  by wrapping   M2-branes  on appropriate  two cycles in such  a proposed  Calabi-Yau  three-fold.      After that,  we  calculate   the  entropy of  these solutions   which takes  a  maximal value  corresponding to $\omega=1$ defining the ordinary projective space  $\mathbb{P}^{4}$.  For  generic values of   $\omega$,  we show that  the non BPS  states  are unstable.   Then, we  conduct  a similar  study  of   five dimensional    black strings.   Concerning    the allowed magnetic  charge  regions  of the  BPS  and non BPS  black stringy solutions   derived  from   M5-branes  on dual divisors,  we calculate the   tension   taking  a    minimal value   for  $\mathbb{P}^{4}$.   By determining    the recombination factor,  we   show  that  the   non-BPS black  string states are  stable in the  allowed  regions in the  magnetic charge  space. 


\end{abstract}

\newpage

\section{Introduction}

Recently,  the    black  objects   in  high energy  theories  have been largely investigated   from   supergravity theories in  arbitrary  dimensional (D) space-times\cite{1,2,3,03,003}.  The corresponding  physical properties of such objects have been dealt with including  the thermodynamic  and  the optical  ones.   Concretely,  black hole behaviors  in   type IIB superstring and  M-theory  have been studied  using different methods \cite{4,5,6,06,7,8,9,10}. In type IIB superstring, for instance,   the thermodynamic  and  the optical   properties of   the black holes in   the  $AdS_5 \times  S^5$   background  have been investigated   using  brane  physics.  Exploiting analytical and numerical methods,   various thermodynamical behaviors  have been approached and examined in arbitrary dimensions by varying the D3-brane number. In particular,  the Hawking-Page transition  has been  treated with and without dark energy  sectors  \cite{7,9}.  These investigations have been   been extended  to M-theory  black holes  with    $AdS_{p+2} \times  S^{11-p-2}$  near horizon  geometries. In this way,  the  black holes can be obtained from   M$p$-branes, where $p=2$ and $ p=5$ producing  solutions  in 4D and 7D, respectively.   Supported by   Even Horizon Telescope  (EHT) observational  findings, the optics   of such   M-theory black holes have been  studied  by dealing with   either  the  shadow  or  the deflection  angle of light rays.    Precisely,  the black hole  shadows  in the  spherical   M-theory compactifications  with M2 and M5 branes   have been examined using the M-brane number  variation \cite{7}.    Using one-dimensional real curves, the 4D   black hole  shadows   have been  studied where various  geometric configurations   have been  obtained. In particular,  the $D$-shapes and the  cardioid  shapes have been  found  for rotating black  holes \cite{9}. Inspired  by such activities,   models supported by  M-theory gravities have been also  approached\cite{11,110}.  Precisely,  the shadows and the deflection angle of  the light rays near  to  the black holes  in the  Starobinsky  Bel-Robinson  gravity    have been  discussed    by introducing  a new stringy parameter $\beta$\cite{12}. It has been shown that such a parameter modifies the  thermodynamics and  optics of the ordinary black holes \cite{12,13,14}.

More recently,  supersymmetric compactifications  have been  exploited to approach  the  BPS and  non BPS  black  objects   from type II superstrings and M-theory \cite{15}.    These studies have been encouraged by the fact that   supersymmetry is broken in lower dimensional realistic  space-times. In 5D,  the  BPS and  non BPS black  objects obtained from M-theory on various Calabi-Yau (CY) three-folds,  with  Kahler moduli parameters,   have  been  investigated  by  calculating  the relevant quantities such as the entropy and the    string  tension using   the  4D   $N=2$ supersymmetry formalism combined with the   attractor mechanism  \cite{153,154,155,156}.  
 To  build  certain  BPS and  non BPS  states,   M2 and M5 branes wrapping two and  four-cycles in CY  geometries using   projective hypersurfaces with two Kahler parameters  have been exploited using techniques developed in \cite{15,16}. In this way, the  relevant quantities of the  black holes  and  the black strings have been computed  and analyzed  to inspect the corresponding  stability behaviors.   Later soon,  these works  have been  generalized  to CY models with three Kahler parameters using  numerical computations via a general treatment  \cite{17}.  
 
In this work,  we  would like  to contribute to these activities by      reconsidering the study of   certain behaviors  of  5D   black    objects      from  the  compactification of M-theory 
on  	 a CY three-fold (CY$^3$) via the     $N = 2$  supergravity formalism.   First, we construct  such a CY$^3$ as   the intersection of two homogeneous polynomials of degrees $  (\omega+2,1)$ and $ (2,1) $ in  a product of  two weighted projective spaces  given by  $ \mathbb{WP}^{4}(\omega,1,1,1,1) \times\mathbb{P}^{1}$.    By  determining  the allowed electric charge  regions  of the  BPS  and non BPS  black hole  solutions   by wrapping  an M2-brane  on appropriate  two cycles in such a  CY$^3$, we   calculate   their   entropies as a function of $\omega$    taking   a  maximal value  associated with the ordinary projective space  $\mathbb{P}^{4}$.  For  generic values of   $\omega$,  we  find  that the   non BPS  states   are unstable matching perfectly with the results of   \cite{15}.   Then, we  provide  a similar  study  of  5D  black string solutions.  For the   the allowed magnetic  charge  regions  of the  BPS  and non BPS  black strings,  we  determine the   tension   involving a    minimal value    for   $\mathbb{P}^{4}$.   By  computing     the recombination factor,  we   show  that  the   non-BPS black  string states are  stable in the  allowed   regions of the magnetic charge space.

The organization of this work is as follows. In section 2, we review  briefly  5D black  objects from M-theory on CY three-folds.   In section 3, we study  the  stability   behaviors of  5D BPS and non-BPS black holes from M2-branes.   In section 4, we  investigate  the 5D BPS and   non-BPS black  strings  from M5-branes.   In the last section, we   present our concluding remarks.

\section{5D black objects in M-theory on Calabi-Yau manifolds}
In this work, we reconsider the study of   behaviors of 5D  black objects   derived from M-theory  CY compactifications.  Before going  ahead, we present  the  relevant concepts which  will be exploited in the  present discussion of  M-theory black object  physical properties. To start,  its is recalled that    a CY three-fold is a compact Kahler  complex  geometry   involving a Ricci-flat metric  with   a SU(3) holonomy  group. The latter has been shown to be  crucial  in string theory revolutions and duality  scenarios\cite{150,151}.  In particular,    the   CY  geometry  has been explored to form a  bridge   between  string theory  in 10D   and  semi-realistic models in 4D with minimal supersymmetric charges  being  $ \frac{1}{4} $ of the initial  ones.  It has been remarked that  a CY three-fold    has  a Hodge diagram  which is   essential  in the determination of the   4D  and 5D  models  with $N=2$ supersymmetry from Type II superstrings and M-theory, respectively.   In connection with  black hole building models,  this manifold  has    been extensively  used   in  the
attractor mechanism\cite{153,154,155}.

 At low energy limits, M-theory   is modeled by  a   $N=1$   supergravity in 11D \cite{20,21,22,220,221}. Omitting    the fermionic  sector,  it contains a  metric $g_{MN}$ and an antisymmetric  three-form  field  $A_{MNP}$, where   $M$, $N$, and $P$ take values from zero to 10.   Besides that,   it involves two  branes known as   M2-branes  and  M5-branes.  M-theory on CY three-folds  lead  to 5D   $N=2$ supergravity  models  with  eight supercharges   involving    vector
multiplets and hypermultiplets.  The  numbers  of such multiplets are fixed by the geometric  Hodge numbers    $h^{1,1}$  and  $h^{2,1}$   counting  the    size and  the shape    parameters, respectively.   It has been remarked that the   shape parameters    are not  relevant in the study of the  black   branes in M-theory  CY  compactifications  by freezing   hypermultiplet fields.   In this way,     the BPS and non BPS   black states are   obtained from  M2 and M5-branes wrapping  on two and four-cycles with  $h^{1,1}$ electric and magnetic charges, respectively.    These solutions can be dealt with by means of a  5D $N=2$  Maxwell-Einstein  supergravity formalism via the following   action
\begin{equation}
S=\frac{1 }{2\kappa_5^2}\int d^5x\left(  R\star \mathbbm{1} -G_{IJ}  dt ^I \wedge  \star dt ^J   -G_{IJ}  F ^I \wedge  \star F^J-  \frac{1 }{6} C_{IJK}   F ^I \wedge  F^J  \wedge  A^K\right)
\end{equation}
where  $I$, $J$, and $K$ take values from  1  to   $h^{1,1}$.   $R$ is the Ricci scalar and   $ t^I$  denote  the size  moduli  associated with  the  Kahler forms $ {\cal J}_I$.     The fields $ F ^I=dA ^I$ are Maxwell tensors corresponding to abelian  vector
multiplets  which can be    obtained from the  $A_{MNP}$   reduction on   the  CY three-folds\cite{23}. The symmetric  tensor  $C_{IJK}$,   which gives  the triple intersection  numbers,   is a crucial  geometric quantity providing   the  normalized volume  of the   CY three-fold   via the relation 
\begin{equation}
{\cal V}= \frac{1 }{3!} C_{IJK}   t ^I  t^J  t^K.
\end{equation}
This quantity is linked  to  the  CY  moduli space metric   via the relation 
\begin{equation}
G^{IJ}= 2\left(-{\cal V} A^{IJ}+ \frac{t ^ I t ^ J}{ 2}\right),
\end{equation}
where   $A^{IJ}$  denotes  the matrix inverse  of   $A_{IJ}= C_{IJK}     t^K.$ Roughly,  the  M-theory compactification on  CY three-folds, in the presence of   M2 and M5 branes wrapping on two and four-cycles,   generates  5D  electric and magnetic charged black   objects,   respectively.   For 5D black holes, the effective potential is expressed  as 
\begin{equation}
\label{Vbh}
V^e_{eff}= G^{IJ} q_Iq_J, 
\end{equation}
where $q_I$ are   the  corresponding  electric charges. To inspect the stability behaviors,  one needs  to determine the potential 
 critical points  by  solving  the constraint 
\begin{equation}
D_IV^e_{eff}= 0 \qquad  D_I= \partial_I-\frac{2}{3{\cal V}} \tau_I,
\end{equation}
where  $ \tau_I$  is given by 
\begin{equation}
\tau_I= \frac{1}{2}  C_{IJK}    t^J t^K ,
\end{equation}
describing the   size of  the $I$-th  divisor  ${\cal D}_I$ in  the  CY three-fold.  These divisors    could  be wrapped by the  M5-brane to provide  the   5D black strings.  

  The  critical points have been exploited to  express certain thermodynamic quantities including the entropy by help of the attractor mechanism \cite{153,154,155}.  For the BPS states, the mass of black holes  is given in terms of  the attractor central charge  value 
\begin{equation}
M=Z_{e}|_{t=t_{c}} 
\end{equation}
where $t_c$  indicates   the  critical value of  $t$. For the non-BPS states, however,  the mass is 
\begin{equation}
M=\sqrt{\dfrac{3}{2}V_{eff}|_{t=t_{c}}},
\end{equation}
leading to 
\begin{equation}
V_{eff}=\frac{2}{3}M^{2}.
\end{equation}
Using the central charge $ Z_{e} $, one gets  
\begin{equation}
S=2\pi\left( \dfrac{Z_{e}|_{t=t_{c}}}{3} \right)^{3/2}.
\end{equation}
The  effective potential  can be used  to express   the non BPS entropy  given as follows 
\begin{equation}
S=2\pi\left( \frac{1}{6}V_{eff}|_{t=t_{c}} \right)^{3/4}.
\end{equation}
More details can be found in  \cite{15}. Similar  techniques have been developed for the black string potential    which reads as  
\begin{equation}
\label{vm}
V^m_{eff}= 4 G_{IJ} p^Ip^J, 
\end{equation}
where   $p^I$  are  now the magnetic charges identified with  the wrapping numbers of   the M5-brane  on  ${\cal D}_I$.  In this way,  the string tension  $T$  is given  in terms  of  the square root of the  magnetic  effective potential calculated  at the critical points obtained by solving  the constraint  $D_I V^m_{eff}=0$.   As suggested   in \cite{15},   the  stability of such 5D black objects   has been discussed    in terms of a ratio called   the recombination factor denoted  by  $R$.  The  solutions are unstable for    $R>1$.  In this situation, the black objects would prefer  to   decay   into the  BPS/non-BPS pairs.  For  $R < 1$,  however,  the black objects are stable  enjoying   the recombination of the brane/anti-brane behaviors. 
\section{ 5D black holes in M-theory CY compactifications}
\begin{figure}[!ht]
  		\begin{center}
						\includegraphics[scale=0.55]{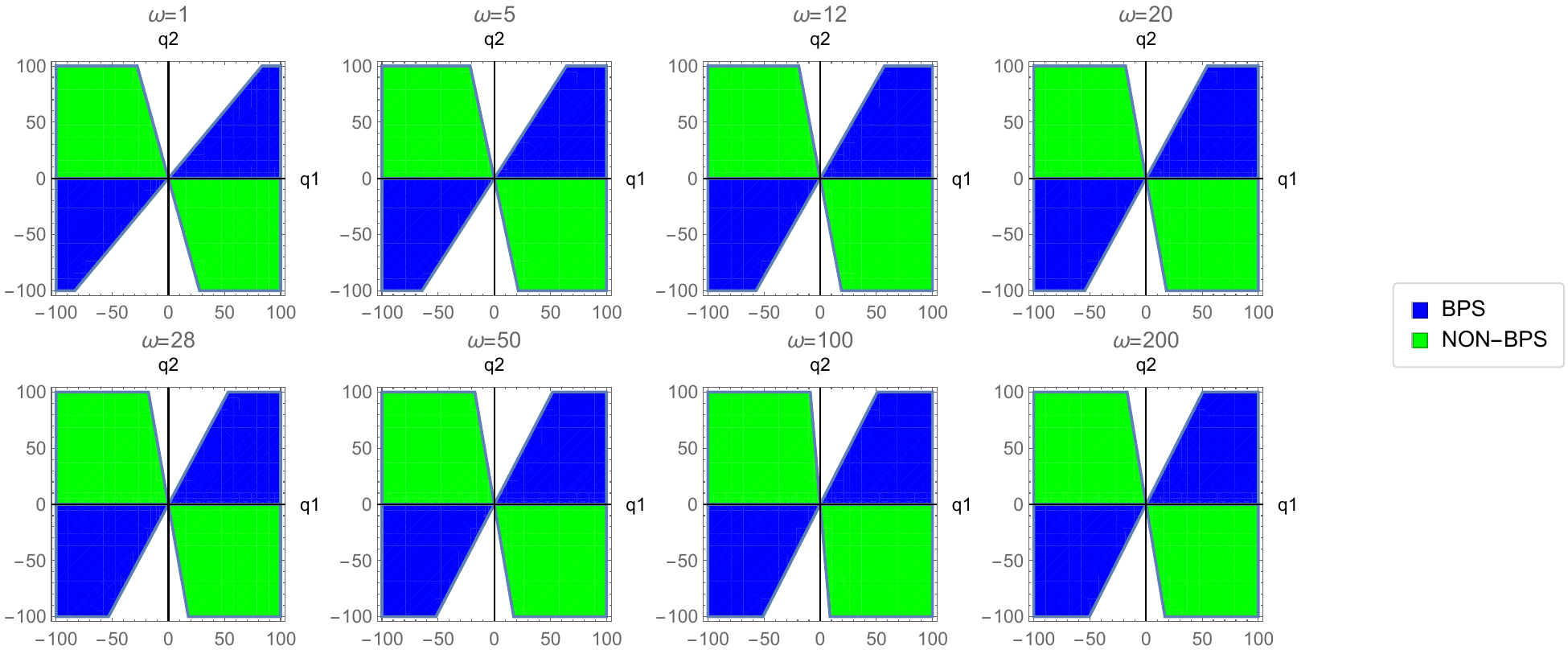} 
	 
\caption{{\it \footnotesize   Electric charge regions for black hole sates. }}
\label{F1}
\end{center}
\end{figure}

In this section, we study  5D black holes from a   special CY$^3$  with two Kahler parameters. The manifold,  that we construct here,  is considered as the intersection of two hypersurfaces in a product of two weighted projective spaces called the ambient geometry given by 
\begin{equation}
\mathcal{A}=\mathbb{W}\mathbb{P}^{4}(\omega_{1},\omega_{2},\omega_{3},\omega_{4},\omega_{5})\times\mathbb{W}\mathbb{P}^{1}(\omega'_{1},\omega'_{2}),
\end{equation}
where $ \omega_{i} $ and  $ \omega'_{i} $   are  natural numbers.   It is recalled that a   $n$-dimensional weighted projective space   $ \mathbb{WP}^{n}$   can be used to extend  the
notion  of  the  ordinary projective space  $ \mathbb{P}^{n}$. Indeed,  $ \mathbb{WP}^{n}$   is defined by  considering the following  identification 
\begin{equation}
z_i \sim \lambda^{\omega_i }z_i, \qquad i=1,\ldots, n+1, 
\end{equation}
where $(z_1,\ldots, z_{n+1})$ are  the  homogeneous coordinates.  $\omega_i$ are called  the weights and  $\lambda$  is a nonzero  complex number. Since the weights are not relevant for the  1-dimensional complex projective space $\mathbb{P}^{1}$,  one  should consider the following ambient space
\begin{equation}
\mathcal{A}=\mathbb{WP}^{4}(\omega_{1},\omega_{2},\omega_{3},\omega_{4},\omega_{5})\times\mathbb{P}^{1}.
\end{equation}
 In this way,   CY$^3$  is associated with a matrix configuration which takes the following form 
\begin{equation}
\left[\begin{array}{cc}
\mathbb{W}\mathbb{P}^{4}(\omega_{1},\omega_{2},\omega_{3},\omega_{4},\omega_{5}) \\ 
\mathbb{P}^{1}
\end{array} \right|\left| \begin{array}{cc}
d_{1}^{1}&d_{1}^{2}\\ 
d_{2}^{1}&d_{2}^{2}
\end{array} \right], 
\end{equation}
such that 
\begin{eqnarray}
d_{1}^{1}+d_{1}^{2}&=&\omega_{1}+\omega_{2}+\omega_{3}+\omega_{4}+\omega_{5}\\
d_{2}^{1}+d_{2}^{2}&=&2,
\end{eqnarray}
needed to satisfy the CY condition.  For such a CY  matrix configuration,   CY$^3$  can be built  via   the intersection of two homogeneous polynomials of degrees $  (d_{1}^{1},d_{2}^{1})$ and $ (d_{1}^{2},  d_{2}^{2}) $ in  $ \mathbb{W}\mathbb{P}^{4}(\omega_{1},\omega_{2},\omega_{3},\omega_{4},\omega_{5}) \times\mathbb{P}^{1}$. Instead of elaborating  general  and complex scenarios,  we restrict ourselves to the effect of only one weight on the  5D  black hole behaviors.   The present  simplicity  could  help  one  to stay focused on  the  weight effect.    This  can be  exploited  to  identify  the associated behavior variation. However,   it can be hoped that the generalization could  provide  relevant findings.    This is  beyond the scope of
the present work.  Roughly,  the weight  effect should depend  on the above matrix configuration form. Up to the CY condition, various choices could  be dealt with.    After an examination,  certain choices  generate a  trivial  weight dependance.    However,  other ones lead  to  a relevant  dependance, where  the weight could appear  in   several  physical quantities.  To see that,  we consider   the following  CY$^3 $ matrix configuration
\begin{equation}
\label{matrix}
\left[\begin{array}{cc}
\mathbb{W}\mathbb{P}^{4}(\omega) \\ 
\mathbb{P}^{1}
\end{array} \right|\left| \begin{array}{cc}
\omega+2&2\\ 
1&1
\end{array} \right], 
\end{equation} 
where we have  used 
\begin{eqnarray}
 \omega_{1}&=&\omega, \quad 
 \omega_{2}=\omega_{3}=\omega_{4}=\omega_{5}=1\\
d_{2}^{1}&=&d_{2}^{2}=1, \quad d_{1}^{1}=\omega+2,   \quad 
d_{1}^{2}=2.
\end{eqnarray}
In this situation,   CY$^3$   is   the intersection of two homogeneous polynomials of degrees $  (\omega+2,1)$ and $ (2,1)$ in  $ \mathbb{WP}^{4}(\omega,1,1,1,1) \times\mathbb{P}^{1}$. 
In this  CY$^3$  configuration, the  5D  black holes can obtained from an M2-brane wrapping a non trivial 2-cycle. The wrapping numbers provide two electric charges $ q_{1} $  and $ q_{2}$ which will be important in the  present section. These two charges generate a relevant ratio $ q=\dfrac{q_{1}}{q_{2}} $ which will be exploited to discuss  the 5D  black hole stability from the M-theory   CY compactification. Before that, we should compute primordial   geometric quantities. In particular, we calculate the normalized volume  of   CY$^3$. This can be obtained by determining the intersection numbers $ C_{IJK} $. Using the method of \cite{24,25}, we find
\begin{equation}
 C_{111} =\omega+4, \qquad  C_{112} =2(\omega+2).
\end{equation}
This  provides  the   normalized  volume expression  of  the proposed   CY$^3 $ given by 
\begin{equation}
{\cal V}=\frac{1}{6} t_{1}^{2}((\omega+4)t_{1}+6(\omega+2)t_{2}).
\end{equation}
Exploiting  the constraint $ {\cal V}=1 $,  we obtain  the effective potential of  the 5D  black holes 
 \begin{equation}
V^e_{eff}= t_{1}^{2}q_{1}^{2}+\dfrac{(4+\omega)t_{1}^{2}}{6(2+\omega)}q_{1}q_{2}+\dfrac{(4+\omega)t_{1}^{2}+8(8+6\omega+\omega^{2})t_{1}t_{2}+24(2+\omega)^{2}t_{2}^{2}}{12(2+\omega)^{2}}q_{2}^{2},
\end{equation}
which is a quadratic polynomial in the electric  charge space $ (q_{1},q_{2}) $. To examine the  5D  black hole behaviors derived   from the proposed    CY$^3 $,  we need to solve the  constraint  $ {D}_{I}V^e_{eff}=0 $.
A  calculation  shows that  we  have the following algebraic equation   
 \begin{equation}
 \label{bpsbh}
 \left(4(2+\omega)q_2-(2(2+\omega)q_1-(4+\omega)q_2)x \right) \left( 12(2+\omega)q_2+(6(2+\omega)q_1+( 4+\omega)q_2)x \right)=0,
\end{equation}
where one has used a local variable $x= \frac{t_1}{t_2}.$ 
It has been remarked  that there are two solutions  which could be given  in terms of the weight  parameter $ \omega $. The latter can be considered as a relevant parameter in the present investigation. In what follows,  the BPS and non-BPS solutions will be dealt with in terms of such a parameter.  For the BPS solutions, we find 
\begin{equation}
x= \dfrac{4(2+\omega)q_2}{2(2+\omega)q_1+(4+\omega)q_2}.
\end{equation} 
The positive values  of the local variable   $ x $   generate two allowed regions in the electric  charge   space which are 
\begin{equation}
\left\{ q_2 < 0 \,\,\, \mbox{and}\,\,\, q_1<\dfrac{(4+\omega)q_2}{2(2+\omega)}\right\},  \,\,\, \,\,\,\left\{q_2 > 0 \,\,\,  \mbox{and},\,\, q_1>\dfrac{(4+\omega)q_2}{2(2+\omega)}\right\}. 
\end{equation}
However, the non-BPS black hole states  correspond to  the  second solution 
\begin{equation}
x= -\dfrac{12(2+\omega)q_2}{6(2+\omega)q_1+(4+\omega)q_2}.
\end{equation} 
This generates  two allowed possible charge   regions
\begin{equation}
\left\{ q_2 < 0 \,\,\,  \mbox{and}\,\,\, q_1<-\dfrac{(4+\omega)q_2}{6(2+\omega)q_1+(4+\omega)q_2}\right\},  \,\,\, \,\,\,\left\{q_2 > 0 \,\,\,  \mbox{and}\,\,\, q_1>-\dfrac{(4+\omega)q_2}{6(2+\omega)q_1+(4+\omega)q_2}\right\}.
\end{equation}
The electric  charge  regions of  the BPS and non-BPS black holes by varying  $\omega$ are illustrated in Fig.(\ref{F1}).
The colored  regions  describe   the existence of the  large BPS  and  non-BPS  black holes  with  non zero electric charges.  It has been remarked that the size of such  regions  depends on   $\omega$. In the non colored regions, the   black hole solutions  are  not allowed.    The size of  the regions  which do not correspond to large black holes  decreases with  the weight   $\omega$.

\begin{figure}[!ht]
		\begin{center}
						\includegraphics[scale=0.75]{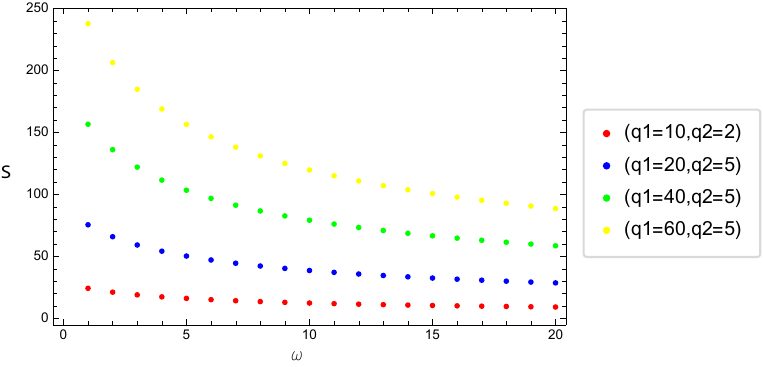} 
	 
\caption{{\it \footnotesize   BPS  entropy behaviors  in terms of    $\omega$.}}
\label{F2}
\end{center}
\end{figure}

The entropy of 5D  black holes can be determined  by considering the constraint $ \mathcal{V}=1 $. The  calculation  gives 
\begin{equation}
S_{BPS}=  \frac{\pi}{6} \sqrt{\dfrac{q_2}{(2+\omega)^{3}}}\,\,(6q_1(2+\omega)-q_2(4+\omega)).
\end{equation} 
This computation is  presented in Fig.(\ref{F2}).
\ 
It follows from this figure that the entropy decreases with the weight  $\omega$. In fact, it starts from a maximal value corresponding to the ordinary projective space   $ \mathbb{P}^{4}$.   This maximal value increases with the  electric charge $q_1$. For large values of   $\omega$, the entropy vanishes.
Calculations  give the expression of the   entropy of the  non-BPS   black  holes which reads as 
 \begin{equation}
S_{non-BPS}=  \frac{\pi}{6} \sqrt{\dfrac{|q_2|}{(2+\omega)^{3}}}\,\,|6q_1(2+\omega)-q_2(4+\omega)|
\end{equation}
exhibiting  similar behaviors with respect to the   $\omega$ dependance.  The   relevant difference appears in the allowed electric charge  regions of the associated black hole moduli space. 

\begin{figure}[!ht]
  		\begin{center}
						\includegraphics[scale=1]{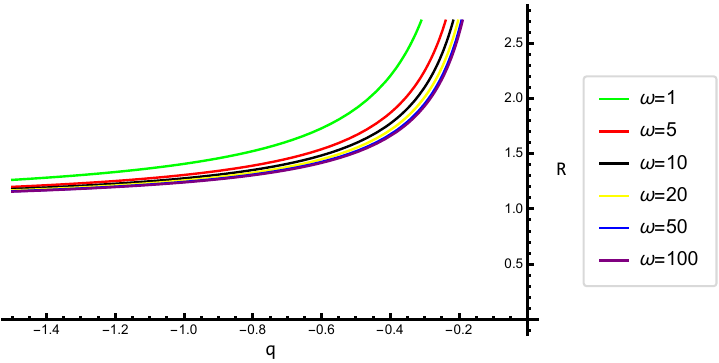} 
	 
\caption{{\it \footnotesize  Recombination factor  of   the non-BPS black hole states by varying $\omega.$}}
\label{RBH}
\end{center}
\end{figure}

The   stability behaviors  of  the non-BPS black holes can be approached   via the recombination factor   $R$  which is  firstly introduced in \cite{15}.   It has been suggested that   $R$ is  the  ratio of the non-BPS black hole mass  to  the M2-brane mass   wrapping  the associated  piecewise calibrated two-cycle.   For  $R > 1$,  the non-BPS black hole is unstable.  The non-BPS states would  prefer  to decay into the associated 
BPS and anti-BPS constituent  states.    For $R < 1$, however,    the constituent BPS-anti-BPS pairs  could  recombine  to provide     stable non-BPS states  in the black hole  spectrum.
   Using the result of  \cite{15},  we   can  determine the recombination factor $ R $ by  considering  the ratio of the non-BPS black hole mass $ V_{c} $ to the  mass of M2-brane  wrapping the associated piecewise calibrated two-cycle $  V_{c^{\cup}} $.   Exploiting the critical  value $t_{c}$,    it  can be expressed as follows 
\begin{equation}
R=\dfrac{V_{c} }{V_{c^{\cup}}}|_{t=t_{c}}
\end{equation} 
where $ V_{c} $  represents the  the  non-BPS black hole mass  depending  on the   effective potential    square via  the  relation  $V_{c}=\sqrt{\frac{3}{2}V_{eff}}$.  After computations, we obtain  
\begin{equation}
V_{c} =\frac{3q_2 t_2 (-6 (\omega+2 )q_1+ (\omega +4)q_2)}{6( \omega +2 )q_1+(\omega +) q_2}.
\end{equation}
The  mass of  an M2-brane  wrapping the associated piecewise calibrated two-cycle $  V_{c^{\cup}} $   reads as
\begin{equation}
V_{c^{\cup}}=t_1|q_1|+t2|q_2|=t_2(|q_1|x+|q_2|).
\end{equation}
Using the critical solution of  the non BPS objects  and  the  first allowed charge region,  this  quantity is found to  be 
\begin{equation}
V_{c^{\cup}}=t_2q_2 \frac{-18(\omega +2)q_1- (\omega +4)q_2}{6 q_1 (\omega +2)+q_2 (\omega +4)}.
\end{equation}
Combining these relations,     we get the  recombination factor 
\begin{equation}
R= 1-\dfrac{4(\omega+4)}{18(\omega+2)q+(\omega+4)},
\end{equation}
where one has  used  the electric  charge ratio $q=\frac{q_1}{q_2}$  for mixed signs of  $q_1$ and $q_2$.  A close examination shows that this ratio is constrained by   $-\frac{\omega+4}{6(\omega+2)}<q<-\frac{\omega+4}{18(\omega+2)}$. For  this range,   we illustrate  the black hole   recombination factor 
 in  Fig.(\ref{RBH}).

It follows from this figure that, for  generic values of  $ \omega$ with  a negative charge ratio  $q$ as required by the above regions,   the non-BPS black hole states are unstable preferring  to decay into the corresponding
BPS and anti-BPS  brane objects.   This matches perfectly with   the  recent results   suggesting that all  the non-BPS black hole states  are unstable   \cite{15}. 

\section{Black string behaviors  in M-theory on CY three-folds}
\begin{figure}[!ht]
  		\begin{center}
						\includegraphics[scale=0.5]{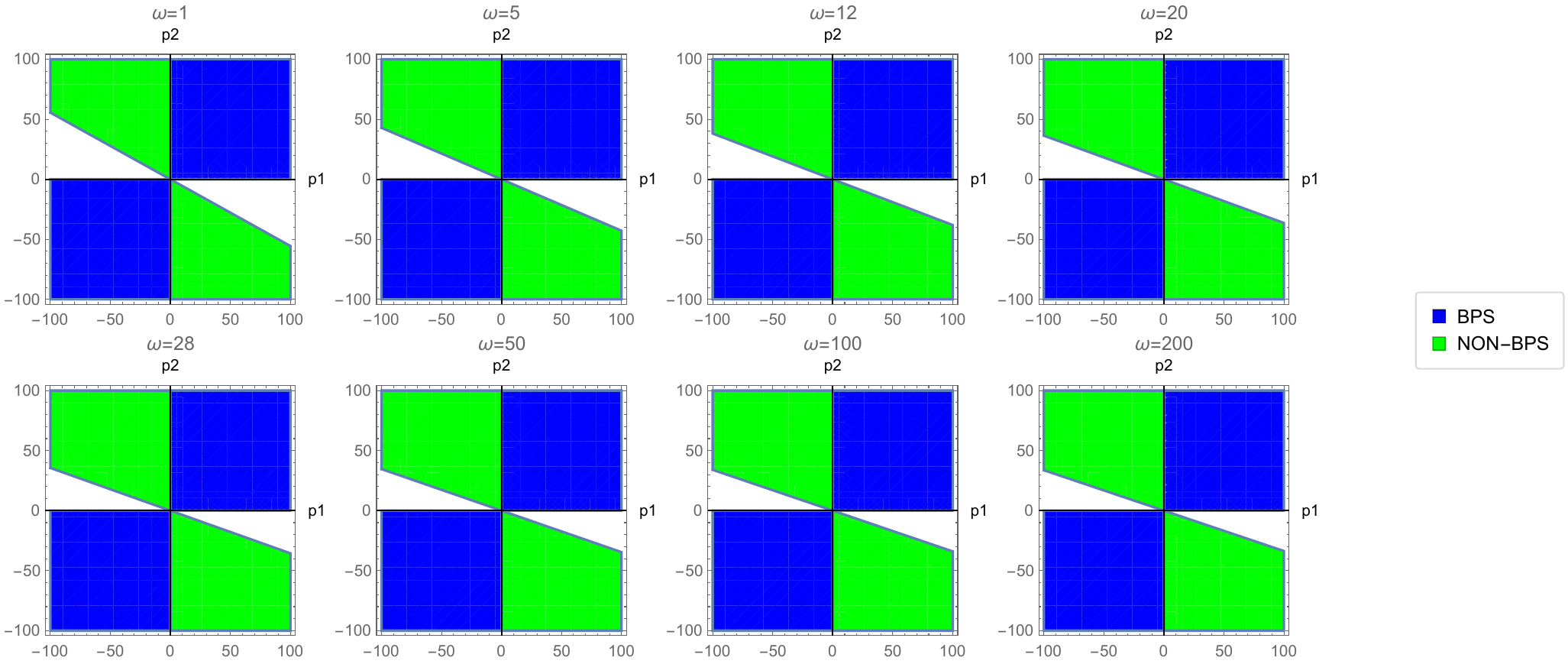} 
	 
\caption{{\it \footnotesize  Allowed charge regions of BPS and non-BPS black strings by varying $\omega$}}
\label{F5}
\end{center}
\end{figure}

  Here, we consider the  5D  black string behaviors in the proposed CY$^3$. These black  branes are obtained by wrapping a dual M5-brane on a generic divisor $ {\cal D} $ which provide  the BPS and non-BPS states  with two  magnetic charges  $ p_1 $ and $ p_2 $.  For such  a CY$^3$, we find that   the  effective  potential  can be  expressed as 
  \begin{equation}
 V^m_{eff}=v_{11}p_1^{2}+ v_{12}p_1p_2+v_{22}p_2^{2}
\end{equation}
  where one has 
  \begin{eqnarray}
v_{11}&=&\frac{1}{6}t_1^{2}\left( 24(2+\omega)^{2}t_2^{2}+8t_1t_2(8+6\omega+\omega^{2})+t_1^{2}(4+\omega)^{2} \right) \nonumber \\
v_{12}&=&
\frac{2}{3}t_1^{4}(2+\omega)(4+\omega)\\
v_{22}&=& 2t_{1}^{4}(2+\omega)^{2}. \nonumber
\end{eqnarray}  
To examine  the 5D  black string behaviors, the constraint $ {D}_{I}V^m_{eff}=0 $ should be solved. This leads  to 
\begin{equation}
( p_1-x p_2 )\left(3 p_1(2+\omega)+x \left(p_1(4+\omega)+3p_2(2+\omega)\right)\right)=0
\end{equation}
where  one has used the local variable  $ x=\dfrac{t_1}{t_2} $. For the BPS black string states, the allowed magnetic charges   corresponding  to the positive values of  such a local variable  is required by 
\begin{equation}
x= \dfrac{p_1}{p_2}.
\end{equation} 
For these BPS  stringy solutions, we have two possible regions in the magnetic charge space
\begin{equation}
\left\{ p_1 < 0 \,\,\, \mbox{and}\,\,\, p_2<0\right\},  \,\,\, \,\,\,\left\{p_1 > 0 \,\,\, \mbox{and}\,\,\, p_2>0\right\}. 
\end{equation}
The non-BPS black string states are associated with  the second solution of the above algebraic equation in the magnetic charge space 
\begin{equation}
x= -\dfrac{3(2+\omega)p_2}{(4+\omega)p_1+3(2+\omega)p_2}.
\end{equation} 
Similarly, the positive values of this local variable of  the  CY moduli space provide two possible regions given by 
\begin{equation}
\left\{ p_2 < 0 \,\,\, \mbox{and}\,\,\, p_1<-\frac{3(2+\omega)p_2}{4+\omega}\right\},  \,\,\, \,\,\,\left\{p_2 > 0 \,\,\, \mbox{and}\,\,\, p_1>-\frac{3(2+\omega)p_2}{4+\omega}\right\}.
\end{equation}
To see  the  corresponding  behaviors, the  allowed magnetic  charge regions  for  the BPS and non-BPS black strings are depicted  in  Fig.(\ref{F5}).

\begin{figure}[!ht]
  		\begin{center}
						\includegraphics[scale=0.8]{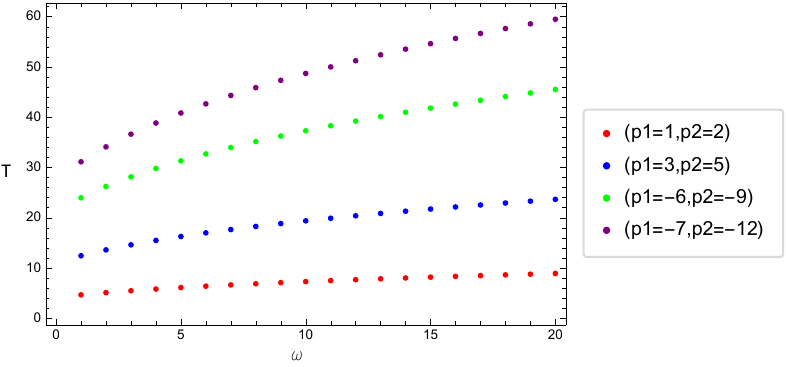} 
	 
\caption{{\it \footnotesize  BPS  brane  tension by varying $\omega$}}
\label{F6}
\end{center}
\end{figure}

\begin{figure}[!ht]
  		\begin{center}
						\includegraphics[scale=0.8]{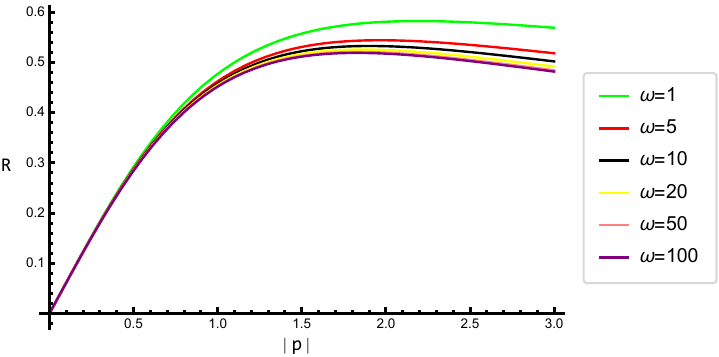} 
	 
\caption{{\it \footnotesize    Recombination factor  of   the non-BPS black string states as  a function of $|p|$  by varying  $\omega$.}}
\label{RBS}
\end{center}
\end{figure}

It has been observed from  this figure that the allowed regions of the  non-BPS black strings depends slightly on   $\omega$.   Precisely, it has been remarked  that the size regions  increases with $\omega$.   There are also some  regions which  are not associated with  black strings. Their sizes decrease  with $\omega$.    For  the  BPS    solutions, the tension  of the string is found to be 
 \begin{equation}
T_{BPS}= \left(   \sqrt{6}p_{1}^{2}( (4+\omega)p_{1}+6(2+\omega)p_{2}   ) \right)^{\frac{1}{3}}.
\end{equation}
The variation  of this function is illustrated in   Fig.(\ref{F6}).
It has been remarked that  $T_{BPS}$ augments with  $\omega$ starting from  a minimal value associated  with    $ \mathbb{P}^{4}$.    Fixing $\omega$,  $T_{BPS}$  increases with   $|p_1|$.   Similar  calculations  can   provide  the expression  of the  BPS  black string  tension. Indeed, it is given by 
 \begin{equation}
T_{Non-BPS}= \left(   |\sqrt{6}p_{1}^{2}( (4+\omega)p_{1}+6(2+\omega)p_{2}   )| \right)^{\frac{1}{3}}.
\end{equation}
 To examine the  stability behaviors, we compute the  recombination factor  $R$  for  such  black strings. The latter  is given  by the ratio of the black string tension to that of the minimal  size of  the  associated  piecewise calibrated divisor.  In terms of a   magnetic charge ratio $p=\frac{p_1}{p_2}$,   we  can calculate such a  recombination factor $R$.  
     This   can  be determined  by the  ratio of the non-BPS black string tension $ T $ to the volume $ V_{D^{\cup}} $ being  the minimum volume piecewise calibrated representative of the class $ [D] $
 \begin{equation} 
 D=p_{1} {\cal J}_1+p_{2}{\cal J}_2
 \end{equation}
  where  $ {\cal J}_1$ and ${\cal J}_2$  denote the Kahler forms of   $\mathbb{WP}^{4}(\omega,1,1,1,1)$ and  $
\mathbb{P}^{1}$, respectively. In this way,  the recombination  factor  reads as 
 \begin{equation}
R=\frac{T}{ V_{D^{\cup}}} 
\end{equation} 
where one has used $ V_{D^{\cup}}=A_{1}|p_{1}|+A_{1}|p_{2}|$ and  $ A_{I}=C_{IJK}t^{J}t^{K}=2\tau_{I} $ describing the size of the divisor in CY$^{3}$.  Concretely,  we get 
 \begin{equation}
R= \sqrt{\dfrac{3}{2}}\left(   \dfrac{(4+\omega)p^{2}+6(2+\omega)p}{6(2+\omega)p^{2}+(4+\omega)p+12(2+\omega)}        \right)
\end{equation}
where the  charge ratio $p$ is constrained  by     $  -\frac{3(2+\omega)}{4+\omega}<p<0$ as required by the allowed magnetic charge regions.   For the proposed   CY$^3$,   the possible range of  the magnetic charges  could be extended  to    $  -3<p<0$ by considering  large weight values.  In this range,  the recombination factor  of   the non-BPS black string states  is illustrated in   Fig.(\ref{RBS}) by varying $\omega$. 

For generic values of $\omega$,  it  follows from this figure  that $R<1$. This shows   that   the   non-BPS black  string states are  stable in the  allowed  magnetic charge regions.  In this way,   they  enjoy  the recombination of the brane/anti-brane behaviors.

\section{Conclusions and discussions}
Using    $N = 2$  supergravity formalism,  we have investigated  certain physical  behaviors  of 5D  black    objects       via   the  compactification of M-theory 
on  	 a special  Calabi-Yau three-fold.    We have first   built  such a  manifold   using the techniques of  the  projective spaces. In particular, the manifold has been  considered as     the intersection of two homogeneous polynomials of degrees $  (\omega+2,1)$ and $ (2,1) $ in  a product of weighted projective spaces given by  $ \mathbb{WP}^{4}(\omega,1,1,1,1) \times\mathbb{P}^{1}$  encoded in  a matrix CY configuration given by Eq.(\ref{matrix}).  The critical points obtained from the effective  potential have been used to     identify   the allowed electric charge  regions  of the 5D   BPS  and non BPS  black hole  solutions. These  region  states have been derived by wrapping  an  M2-brane  on appropriate  two cycles in such a CY  manifold by varying the weight $\omega$.  Then, we    have calculated   the  entropy of  the obtained    solutions   taking  a  maximal value   corresponding to  the ordinary projective space  $\mathbb{P}^{4}$. For  generic values of   $\omega$,  we   have shown  that  the  non BPS  states,     associated   with the  allowed electric charge  regions,  are unstable by computing the black hole    recombination factor.  This matches  perfectly with the previous works suggesting that all non BPS black holes are unstable \cite{15}.   Finally, we have  elaborated   a similar  study  for  black stringy solutions.   For  the allowed magnetic  charge  regions  of the  BPS  and non BPS  black strings,  we  have determined  the   tension as a function of $\omega$    with  a    minimal value    for   $\mathbb{P}^{4}$.   By  computing     the  stringy recombination factor,  we    have revealed   that  the   non-BPS black  string states are  stable in the  allowed  magnetic charge regions. 

This work   comes up with certain open questions.  It could be possible to consider other geometries with non trivial holonomy groups. Other CY  technologies could be exploited to unveil extra data associated with  such activities    by considering  more than one weight. Mirror symmetry could find a place in such black hole studies.  Motivated  by our recent works on optical behaviors of  black holes, the shadow and the deflection angle of light  near to  the obtained solutions could be approached using the developed techniques in arbitrary dimensions.   We hope come back to such questions in future investigations.

\textbf{Acknowledgement}:   One of the authors  (AB)   would like to thank   N. Askour,   S. E. Baddis,  H. Belmahi, M. Benali,  H. El Moumni and Y. Sekhmani  for collaborations on related topics.   He is     grateful to his  family  for support, especially  his mother Fatima.  He is wishing  her  a speedy recovery and a healthy, happy,  future.  
    The authors  would like   to thank the editor and the anonymous referee for remarks, suggestions and scientific  helps.

\end{document}